# Rabi-like oscillation of photonic topological valley Hall edge states

Hua Zhong[1,2,3], Yaroslav V. Kartashov[4], Yiqi Zhang[1,2,3,*], Daohong Song[5], Yanpeng Zhang[2], Fuli Li[1], and Zhigang Chen[5,6]

[1]Department of Applied Physics, School of Science, Xi'an Jiaotong University, Xi'an, 710049, China
[2]Key Laboratory for Physical Electronics and Devices of the Ministry of Education & Shaanxi Key Lab of Information Photonic Technique, Xi'an Jiaotong University, Xi'an, Shaanxi 710049, China
[3]Guangdong Xi'an Jiaotong University Academy, Foshan 528300, China
[4]Institute of Spectroscopy, Russian Academy of Sciences, Troitsk, Moscow, 108840, Russia
[5]MOE Key Laboratory of Weak-Light Nonlinear Photonics, TEDA Applied Physics Institute and School of Physics, Nankai University, Tianjin 300457, China
[6]Department of Physics and Astronomy, San Francisco State University, San Francisco, California 94132, USA.
*Corresponding author: zhangyiqi@mail.xjtu.edu.cn

We investigate Rabi-like oscillations of topological valley Hall edge states by introducing two zigzag domain walls in an inversion-symmetry-breaking honeycomb photonic lattice. Such resonant oscillations are stimulated by weak periodic modulation of the lattice depth along the propagation direction that does not affect the overall symmetry and the band topology of the lattice. Oscillations are accompanied by periodic switching between edge states with the same Bloch momentum, but located at different domain walls. Switching period and efficiency are the nonmonotonic functions of the Bloch momentum in the Brillouin zone. We discuss how efficiency of this resonant process depends on detuning of modulation frequency from resonant value. Switching of nonlinear edge states is also briefly discussed. Our work brings about an effective approach to accomplish resonant oscillations of the valley Hall edge states in time-reversal-invariant topological insulators. © 2019 Optical Society of America

*OCIS codes:* (130.5296) Photonic crystal waveguides; (350.5500) Propagation; (160.1245) Artificially engineered materials.

Photonic topological insulators and topological lasers have attracted global attention nowadays [1-8]. In photonics and related areas, topological insulators have been constructed by breaking either the time-reversal symmetry [4,9-13] or the spatial inversion symmetry [14,15]. In the latter case, since the time reversal symmetry is not broken, it is more convenient to realize a feasible topological system that offers topological protection against certain classes of disorder. Among numerous optical lattice structures, the graphene-like honeycomb lattice (HCL) offers a very convenient platform for topological photonics. The HCL has a pair of degenerated but inequivalent Dirac points $\mathbf{K}$ and $\mathbf{K}'$ at the corners of the first Brillouin zone, as has been employed for demonstration of valley pseudospin and valley Landau-Zener-Bloch oscillations [16,17]. The HCL consists of two sublattices, and its inversion symmetry can be broken by setting different refractive indices or different site sizes in two sublattices [14,15]. Breakup of inversion symmetry opens the gap at the Dirac points, leading to a host of fundamental phenomena due to the intriguing valley degree of freedom [18,19]. For instance, the Berry curvature has opposite signs at the $\mathbf{K}$ and $\mathbf{K}'$ valleys, that can be attributed to the effective magnetic field leading to the well-known valley Hall effect [20]. It has been proven both theoretically and experimentally that at the domain walls [14,21,22] there exist robust valley Hall edge states (VHES) topologically protected by symmetry, so they can circumvent sharp corners or obstacles [15] without back-reflection or radiation into the bulk. Inspired by the discoveries in topological electronic systems, a variety of valley-mediated effects including valley polarization, valley pseudospin and vortex states, and valley topological transport have been investigated in photonic and phononic crystals [15,16,23-30], among other systems.

Since the internal scattering of the VHES is inhibited due to topological protection, the energy typically cannot be routed to other locations away from the domain wall. As such, the application of the VHES in switching-based optical devices is limited. Is there a way to overcome this limitation and meanwhile preserve the topological protection? In this Letter, we construct two zigzag domain walls in an inversion-symmetry-breaking HCL, and investigate Rabi-like oscillations of the topological valley Hall edge states residing at these domain walls. The domain walls are introduced by tuning the refractive index difference in different sublattices. Rabi-like oscillations are induced by weak and periodic longitudinal modulation of the HCL [31-33] and they are manifested in periodic switching of the edge states between different domain walls, akin to the Rabi oscillations previously studied with non-topological waveguide arrays [31,34]. We emphasize that our method of imposing a weak modulation along the propagation direction does not affect the topological property and symmetry of the HCL. It is also different from temporal-modulation based edge-state switching previously proposed for polariton topological insulators [33], where the time-reversal symmetry is broken due to the spin-orbit coupling while the inversion symmetry is retained.

The paraxial equation for amplitude $\psi$ of a light beam propagating along the $z$-axis of a longitudinally modulated lattice can be written as:

$$i\frac{\partial\psi}{\partial z}=-\frac{1}{2}(\partial_x^2+\partial_y^2)\psi-\mathcal{R}(x,y)[1+\mu\sin(\omega z)]\psi-g|\psi|^2\psi. \quad (1)$$

Here the transverse $(x,y)$ and longitudinal $(z)$ coordinates are normalized to the characteristic transverse scale $r_0$ and the diffraction length $L_{\text{dif}}=kr_0^2$, respectively; $k=2\pi n_0/\lambda$ is the wavenumber; $n_0$ is the background refractive index; the last term describes a focusing cubic nonlinearity with strength $g>0$; $\mu$ is the depth of the longitudinal modulation and $\omega$ is the modulation frequency. The HCL is described by the function $\mathcal{R}(x,y)=\mathcal{R}_a(x,y)+\mathcal{R}_b(x,y)$, where $\mathcal{R}_a$ and $\mathcal{R}_b$ describe two standard sublattices of the HCL. Each sublattice $\mathcal{R}_{a,b}(x,y)=p_{a,b}\sum_{n,m}\mathcal{Q}(x-x_n,y-y_m)$ is composed of Gaussian waveguides $\mathcal{Q}=\exp(-x^2/a_x^2-y^2/a_y^2)$ with $p_{a,b}$ being the depths of two sublattices and $a_{x,y}$ being waveguide widths. The HCL is truncated in the $y$ direction but extends periodically along the $x$ direction with a periodicity $X=3^{1/2}d$ ($d$ being the lattice constant). We presume that the HCL with two domain walls can be prepared by using, for instance, the femtosecond laser writing technique [4,5] and use the following parameters: $\lambda=1045$ nm, $d=18.5\,\mu\text{m}$, $a_x=4.9\,\mu\text{m}$, $a_y=3.2\,\mu\text{m}$. For transverse scale $r_0=4.9\,\mu\text{m}$, the diffraction length $L_{\text{dif}}\sim 0.2$ mm. The sublattice depths $p_a=0.94$ and $p_b=1.56$ correspond to refractive index changes of $7.5\times 10^{-4}$ and $12.4\times 10^{-4}$ in a real physical system, respectively [4,5].

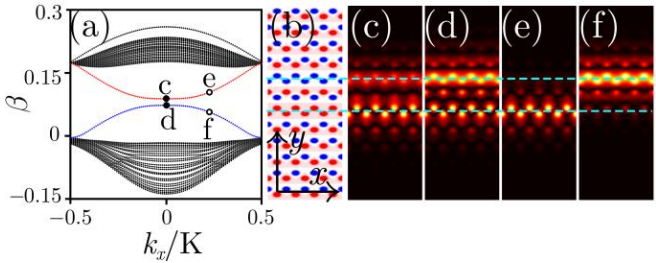

Fig. 1. (a) Band structure and (b) corresponding geometry of the honeycomb lattice. (c)-(f) Stationary valley Hall edge states obtained without longitudinal modulation at (c,d) $k_x=0$, corresponding to the solid circles in (a), and (e, f) $k_x=0.22K$, corresponding to the open circles in (a). Dashed lines in (b)-(f) represent the positions of the two domain walls.

First, we consider linear system without modulation (i.e., $g=0$ and $\mu=0$). Corresponding linear modes are the Bloch states which are periodic along $x$ and localized along $y$; they can be written as $\psi(x,y,z)=u(x,y)\exp(i\beta z+ik_x x)$, where $u(x,y)=u(x+X,y)$, $u(x,y\to\pm\infty)=0$, $k_x$ is the Bloch momentum, and $\beta$ is the propagation constant, which is a periodic function of Bloch momentum $k_x$ with a period $K=2\pi/X$. Plugging the linear mode into Eq. (1), one obtains the eigenproblem $\beta u=[(\partial_x+ik_x)^2+\partial_y^2]u/2+\mathcal{R}u$ which can be solved numerically for the $\beta(k_x)$ spectrum by using the plane-wave expansion method. This spectrum is displayed in Fig. 1(a). The corresponding configuration with two zigzag domain walls (marked by two dashed lines) is shown in Fig. 1(b). Clearly, there are two topological valley edge states as indicated by the red and blue curves in Fig. 1(a), which localize at the bottom and top domain walls in Fig. 1(b), respectively. In Fig. 1(b), the sublattice $\mathcal{R}_a$ ($\mathcal{R}_b$) is marked by blue (red) color. In Figs. 1(c)-1(f), we display topological VHES at two different Bloch momenta. One can observe that depending on Bloch momentum value the edge state may be strongly localized on one domain wall, or it may show appreciable coupling into other domain wall [Figs. 1(c) and 1(d)]. If the Bloch momentum increases, the overlap of the edge states on different domain walls decreases [Figs. 1(e) and 1(f)]. It is apparent that the overlap between different edge states is determined by separation between two domain walls. Since smaller separation allows to achieve faster switching, here we design the lattice such as to have minimal separation between domain walls, see dashed lines in Fig. 1(b).

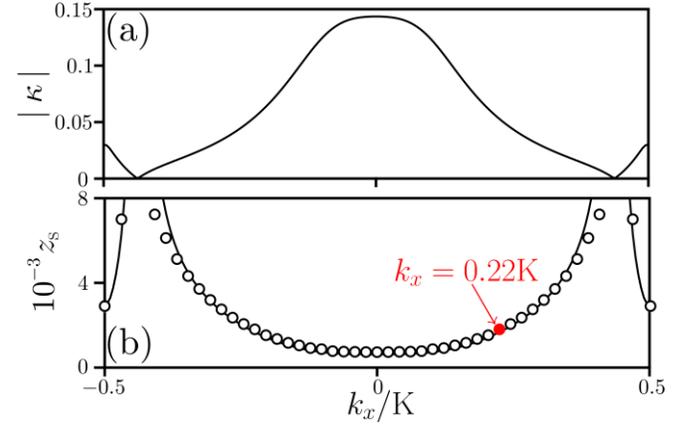

Fig. 2. Coupling coefficient (a) and corresponding switching distance (b) of the two VHES versus Bloch momentum. Solid curves and open circles in (b) are analytical and numerical results, respectively.

In order to verify the effect of weak modulation on evolution of edge states, we use the VHES at the bottom domain wall [red curve in Fig. 1(a)] as the input for Eq. (1) at $\mu\ne 0$. We consider weak modulation, with amplitude $\mu\ll 1$, whose average over one $z$-period $Z=2\pi/\omega$ is zero $\langle\sin(\omega z)\rangle_Z=0$, and analyze the impact of such modulation on the initial modes of unmodulated system. The period $Z$ of modulation is much smaller than actual switching distance, so such fast and weak $z$-modulation should be considered as perturbation that only couples different modes in the initial basis, without notably altering their propagation constants. The modulation frequency is chosen as $\omega=\beta_r-\beta_b+\delta$ with $\delta$ being the modulation frequency detuning and $\beta_{r,b}$ being the propagation constants of the edge states [here the subscript r (b) means red (blue) color in accordance with color of branches in Fig. 1(a)]. $\delta=0$ corresponds to the resonant case when modulation frequency $\omega_0=\beta_r-\beta_b$ amounts to propagation constant difference. Note that we use the following normalizations $\langle u_r,u_r\rangle=\langle u_b,u_b\rangle=1$ and orthogonality $\langle u_r,u_b\rangle=0$. According to the coupled mode theory [31], longitudinal refractive index modulation with near-resonant frequency can couple edge states at two domain walls. The evolution of their amplitudes (that become $z$ dependent) is governed by the equations $\partial c_{r(b)}/\partial z=\mp i\mu\kappa c_{b(r)}\exp(\pm i\delta z)/2$, where $c_{r(b)}$ are slowly varying complex amplitudes of the interface states on different domain walls and the coupling coefficient can be written as:

$$\kappa = \langle u_r, \mathcal{R} u_b \rangle = \int_0^X dx \int_{-\infty}^{+\infty} u_r^* \mathcal{R} u_b \, dy, \quad (2)$$

The corresponding switching distance is defined as $z_s = 2\pi/\mu|\kappa|$ at $\delta = 0$. At $z = z_s/2$ the VHES completes one transition from one domain wall to the other during the Rabi-like oscillation process. Since switching distance is inversely proportional to the modulation depth $\mu$ and coupling coefficient $|\kappa|$, as dictated by coupled-mode equations, the fastest switching occurs at the largest $|\kappa|$. In Fig. 2(a), we show the dependencies of the coupling coefficient on Bloch momentum. This dependence is symmetric with respect to $k_x = 0$, where $|\kappa|$ acquires maximal value. One also finds that this dependence is non-monotonous in the first Brillouin zone: for $|k_x| < 0.43 \text{K}$ coupling coefficient decreases with increase of $|k_x|$, but for $|k_x| > 0.43 \text{K}$ it increases with $|k_x|$. Remarkably, near the points $k_x = \pm 0.43 \text{K}$ the coupling coefficient vanishes. In Fig. 2(b) we show the corresponding switching distance, with the open circles and solid curve being numerical results obtained by direct integration of Eq. (1) and predictions of coupled-mode theory, respectively. One finds that the switching distance predicted by theory perfectly matches switching distance obtained in direct simulations practically for all $k_x$ values, except for points, where coupling constant vanishes. By comparing Figs. 2(a) and 2(b), one finds that the switching distance is indeed inversely proportional to the coupling coefficient (i.e. switching is fastest at $k_x = 0$), and that it diverges when $|\kappa|$ vanishes. The particular value of momentum, where the coupling coefficient vanishes, depends exclusively on the mode shapes $u_r$ and $u_b$, which are in turn defined by the shape and depth of the lattice $\mathcal{R}$.

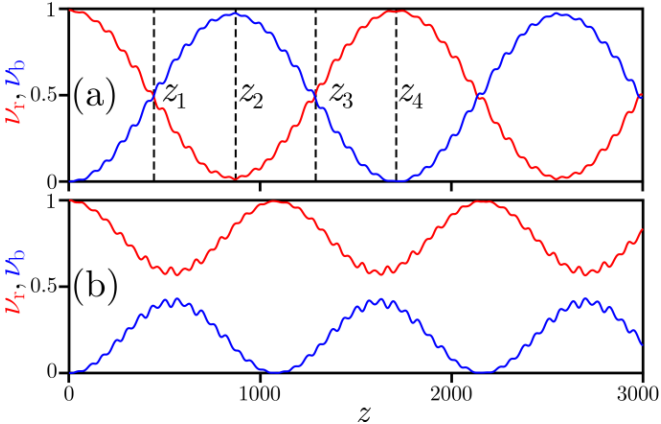

Fig. 3. Evolution of modal weights $\nu_{r,b}$ of the VHES at $k_x = 0.22\text{K}$ illustrating Rabi-like oscillations. (a) Resonant case, $\delta = 0$. Distances $z_i = i z_s / 4$, $i = 1, ..., 4$. (b) Non-resonant case, $\delta = 0.005$. In both cases, the modulation depth is $\mu = 0.07$. The colors of the curves are in accordance with the colors of the VHES in Fig. 1(a).

To illustrate the process of switching, in Fig. 3 we plot the evolution of weights of two edge states during propagation. These weights can be calculated from projections of the field amplitude $\psi$ on the initial stationary edge states: $\nu_{r,b} = |c_{r,b}|^2 = |\langle u_{r,b} e^{ik_x x}, \psi \rangle|^2$. Because radiation into bulk is negligible, the conservation law $\nu_r + \nu_b = 1$ holds, as it is apparent from the coupled-mode equations for amplitudes $c_{r,b}$. The dependencies $\nu_{r,b}(z)$ are displayed in Fig. 3 for both resonant [Fig. 3(a)] and non-resonant [Fig. 3(b)] cases in the linear system at $k_x = 0.22\text{K}$ [see the red dot in Fig. 2(b)]. Notice that fast small-amplitude oscillations in $\nu_{r,b}$ are exactly due to the periodic modulation of the potential with period $Z = 2\pi/\omega$. We would like to mention that it is straightforward to use the maximum value of $\nu_b$ that is labeled as $\nu_b^{max}$, to characterize the efficiency of switching. In Fig. 3(a), the VHES on the bottom domain wall is set as the input. At $z = z_1$, the edge states on the top and bottom domain walls have practically the same weights. Later, at $z = z_2$, the power of the edge state on the top domain wall reaches its maximum – half of Rabi oscillation cycle is completed. After that the reverse energy exchange starts leading to recovery of the initial state of the system at complete cycle, $z = z_4$. The intensity distributions of the VHES at these representative distances during the switching process are illustrated in Fig. 4. One can see that the beam initially concentrated on the bottom domain wall [Fig. 4(a)], switches to the upper domain wall [Fig. 4(c)], and finally returns to the initial position [Fig. 4(e)]. In addition to exact resonance, switching can occur also under non-resonant conditions, for example at $\delta = 0.005$, as shown in Fig. 3(b). In comparison with the resonant case, the switching distance $z_s$ in the non-resonant case is smaller, but the efficiency decreases too, akin to incomplete Rabi oscillations. It should be stressed that since diffraction length is about $0.2 \text{ mm}$ for our parameters, the real-world switching distance constitutes about $17.5 \text{ cm}$ for the edge state at $k_x = 0.22\text{K}$ and it can be even smaller for smaller momentum values, as Fig. 2(b) shows. This confirms feasibility of the experimental observation of this effect in standard samples.

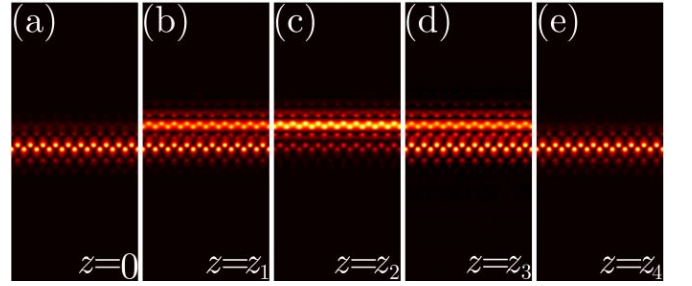

Fig. 4. Transverse intensity patterns of the VHES at representative propagation distances marked in Fig. 3(a) for $k_x = 0.22\text{K}$, at $\mu = 0.07$ and $\delta = 0$. Switching between top and bottom domain walls is evident.

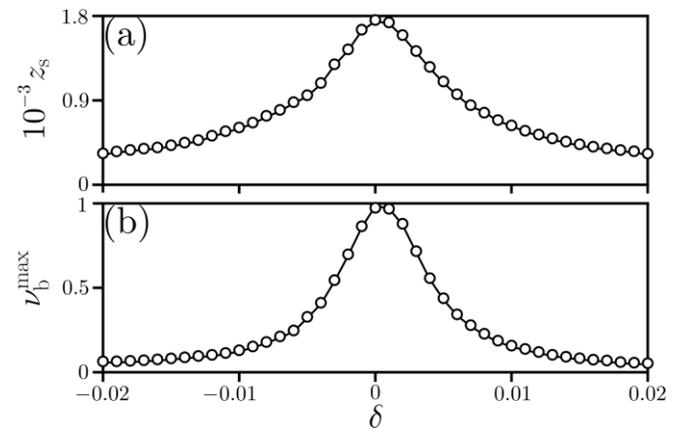

Fig. 5. The switching distance (a) and efficiency (b) of the VHES at $k_x = 0.22\text{K}$ versus the detuning $\delta$ in the linear system (i.e., the nonlinear coefficient $g = 0$) with $\mu = 0.07$.

Since the detuning strongly affects the switching process, as shown in Fig. 3(b), we display in Fig. 5(a) and 5(b) the dependence of switching distance and switching efficiency on detuning $\delta$. One can see that both of them are symmetric about $\delta=0$, and that the switching distance is largest in exact resonance.

At last, we consider switching for nonlinear edge states with $g=1$ in Eq. (1). Such states can be found from the nonlinear problem $\epsilon u = [(\partial_x + ik_x)^2 + \partial_y^2]u/2 + \mathcal{R}u + |u|^2 u$, where $\epsilon$ is the propagation constant of nonlinear VHES, that can be written as $\psi(x,y,z) = u(x,y,z)\exp(i\epsilon z + ik_x x)$. This problem was solved using Newton's iterative method. In Fig. 6(a) we display norm per period $N = \int_{-\infty}^{+\infty} dy \int_{-X/2}^{+X/2} |\psi|^2 dx$ for the nonlinear edge states at two domain walls at $k_x = 0.22K$. Nonlinear edge states bifurcate from their linear counterparts (dashed lines). Their norm grows with $\epsilon$ until propagation constant reaches the border of the gap. We found that the difference of nonlinear propagation constants of two nonlinear edge states with the same norm from Fig. 6(a) gives a very good approximation to the frequency of periodic modulation that causes resonant switching between such nonlinear VHESs. This resonant frequency is depicted in Fig. 6(b). In focusing nonlinear medium with $g=1$ this frequency increases with increase of the norm $N$, because the difference of propagation constants of two states increases with increase of the nonlinearity strength.

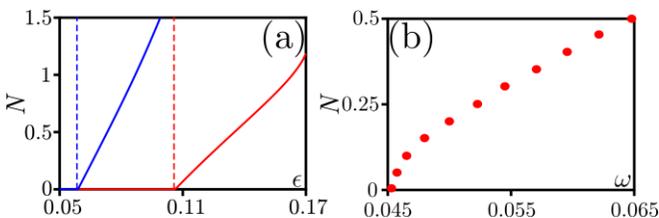

Fig. 6. (a) Norm of nonlinear edge states versus nonlinear propagation constant $\epsilon$. The colors of the curves are in accordance with the colors of the VHES in Fig. 1(a). Dashed lines indicate propagation constants of the linear edge states. (b) Dependence of resonant modulation frequency $\omega$ on norm $N$ of the nonlinear edge states.

In summary, we have theoretically studied the resonant switching of valley Hall edge states. Two domain walls are constructed in an inversion-symmetry-broken HCL by making the two sublattices with different refractive indices. The associated topologically protected valley Hall edge states are obtained, and switching between them is realized when the periodic weak modulation is imposed to the HCL along the propagation direction. The proposed switching method is highly selective due to its resonant nature, it does not change topology of the lattice and is supposed to be robust to weak disorder and inhomogeneities in the array. Our results not only have potential applications for fabrication of switching devices, but also provide a new viewpoint for investigating the valley Hall effect related topological phenomena.

**Funding.** National Key R&D Program of China (2017YFA0303703,2017YFA0303800); RFBR and DFG (18-502-12080); National Natural Science Foundation of China (11534008); Natural Science Foundation of Shaanxi Province (2017JZ019,2016JM6029) and Guangdong Province (2018A0303130057); Ministry of Science and Technology of China (2016YFA0301404); Fundamental Research Funds for the Central Universities (xzy012019038, xzy022019076).